\documentclass[runningheads]{llncs}
\usepackage[T1]{fontenc}
\usepackage{graphicx,verbatim}
\usepackage{amsmath,amssymb}
\usepackage{stmaryrd}
\usepackage{booktabs}
\usepackage{tabularx}
\usepackage{multirow}
\usepackage{multicol}
\usepackage{adjustbox}
\usepackage{bbm}
\usepackage{xcolor,colortbl}
\usepackage{tikz}
\usepackage{caption}
\usepackage{subfig}
\usepackage{stackengine}
\usepackage{makecell}
\usepackage{hyperref}
\usepackage{color}
\usepackage{subcaption}
\usepackage{marvosym}
\newcommand{\EnvelopeAuthor}{\textsuperscript{\Letter}}

\begin{document}
%
% \title{MCM: Mamba-based Cardiac Motion Tracking in MRI}
% \title{MCM: Mamba-based Cardiac Motion Tracking using Sequential MRI Images}
\title{MCM: Mamba-based Cardiac Motion Tracking using Sequential Images in MRI}

\titlerunning{Mamba-based Cardiac Motion Tracking}
%
\begin{comment}  %% Removed for anonymized MICCAI 2025 submission
\author{First Author\inst{1}\orcidID{0000-1111-2222-3333} \and
Second Author\inst{2,3}\orcidID{1111-2222-3333-4444} \and
Third Author\inst{3}\orcidID{2222--3333-4444-5555}}
%
\authorrunning{F. Author et al.}
\institute{Princeton University, Princeton NJ 08544, USA \and
Springer Heidelberg, Tiergartenstr. 17, 69121 Heidelberg, Germany
\email{lncs@springer.com}\\
\url{http://www.springer.com/gp/computer-science/lncs} \and
ABC Institute, Rupert-Karls-University Heidelberg, Heidelberg, Germany\\
\email{\{abc,lncs\}@uni-heidelberg.de}}

\end{comment}

% \author{Paper ID 8}  %% Added for anonymized MICCAI 2025 submission
% \authorrunning{***}
% % \institute{Anonymized Affiliations \\
% %     \email{email@anonymized.com}}
% \institute{*** }

\author{Jiahui Yin\inst{1} \and
Xinxing Cheng\inst{1} \and
Jinming Duan\inst{1,2,3} \and
Yan Pang\inst{4} \and \\
Declan O’Regan\inst{5} \and
Hadrien Reynaud\inst{6,7} \and
Qingjie Meng\inst{1,7}\EnvelopeAuthor}

\authorrunning{J. Yin et al.}
\institute{
School of Computer Science, University of Birmingham, Birmingham, UK\\
\email{\{jxy427, m.qingjie\}@bham.ac.uk} \and
Division of Informatics, Imaging and Data Sciences, University of Manchester, Manchester, UK \and
Centre for Computational Imaging and Modelling in Medicine, University of Manchester, Manchester, UK \and
Guangdong Provincial Key Laboratory of Computer Vision and Virtual Reality Technology, Shenzhen Institute of Advanced Technology, Chinese Academy of Sciences, Shenzhen, China \and
MRC Laboratory of Medical Sciences, Imperial College London, London, UK \and
UKRI CDT in AI for Healthcare, Imperial College London, London, UK \and
Department of Computing, Imperial College London, London, UK}

\maketitle              % typeset the header of the contribution
\begin{abstract}
{
Myocardial motion tracking is important for assessing cardiac function and diagnosing cardiovascular diseases, for which cine cardiac magnetic resonance (CMR) has been established as the gold standard imaging modality. Many existing methods learn motion from single image pairs consisting of a reference frame and a randomly selected target frame from the cardiac cycle. However, these methods overlook the continuous nature of cardiac motion and often yield inconsistent and non-smooth motion estimations. In this work, we propose a novel Mamba-based cardiac motion tracking network (MCM) that explicitly incorporates target image sequence from the cardiac cycle to achieve smooth and temporally consistent motion tracking. By developing a bi-directional Mamba block equipped with a bi-directional scanning mechanism, our method facilitates the estimation of plausible deformation fields. With our proposed motion decoder that integrates motion information from frames adjacent to the target frame, our method further enhances temporal coherence. Moreover, by taking advantage of Mamba’s structured state-space formulation, the proposed method learns the continuous dynamics of the myocardium from sequential images without increasing computational complexity. We evaluate the proposed method on two public datasets. The experimental results demonstrate that the proposed method quantitatively and qualitatively outperforms both conventional and state-of-the-art learning-based cardiac motion tracking methods. The code is available at https://github.com/yjh-0104/MCM.

}

% Authors must provide keywords and are not allowed to remove this Keyword section.

\end{abstract}

\section{Introduction}
Left ventricular (LV) myocardial motion tracking enables the assessment of LV function spatially and temporally~\cite{Puyol2019,inacio2023}. This facilitates the early and accurate detection of LV dysfunction and myocardial diseases~\cite{Ibrahim2011,Claus2015,Bello2019}. Cine cardiac magnetic resonance (CMR) imaging is widely employed in myocardial motion tracking, as it provides high-resolution 2D image sequences that capture detailed structural and functional information of the heart. Recent advancements in deep learning have been leveraged for cardiac motion estimation in CMR images~\cite{Ye2021,YuH2020,Meng2022_mulvimotion,QIN2023102682,Meng2023_deepmesh,Meng2022_miccai,Beetz2024}. Many methods train neural networks to learn the motion between a reference frame and a randomly selected target frame from the cardiac cycle. However, by focusing on isolated target frame, these approaches overlook the continuous nature of cardiac motion. This often results in motion estimations that lack consistency and smoothness. Although incorporating the entire sequence of images could address these issues, it would introduce significant memory and computational challenges.

In this work, we propose a novel Mamba-based network that utilizes a sequence of target frames for improved myocardial motion tracking. Our method explicitly incorporates neighboring frames around the target frame to estimate the motion between the reference and the target frame, which enables the estimation of consistent and smooth motion fields. By leveraging Mamba’s structured state-space formulation, the proposed approach effectively learns the continuous dynamics of the myocardium from the target image sequences with no significant increase in computational complexity. Moreover, our method integrates spatiotemporal information from both forward and backward directions, facilitating the estimation of plausible deformation fields during motion tracking.

\noindent\textbf{Contributions:}
(1) We propose an end-to-end trainable Mamba-based cardiac motion tracking network (MCM) that leverages sequential images to achieve smooth and consistent myocardial motion estimation without incurring significant computational overhead. 
% (2) We introduce a bi-directional scanning mechanism in Mamba block to learn spatiotemporal information from both forward and backward direction, enabling the estimation of plausible
% deformation fields. 
(2) We introduce bi-directional Mamba blocks to extract deformation features at multiple scales. Each block incorporates a novel bi-directional scanning mechanism that captures spatiotemporal information in both forward and backward directions, facilitating the estimation of plausible deformation fields. 
% (3) We develop a motion decoder to estimate a temporal consistent motion field by fusing deformation features from neighboring frames at various scales. 
(3) We develop a motion decoder that estimates motion fields by fusing deformation features across multiple scales, incorporating a novel dual-path fusion head to enhance the temporal consistency of motion estimation.

% 4) Comprehensive experiments on two public datasets demonstrate that our method achieves state-of-the-art performance, outperforming existing approaches.

% \noindent\textbf{Contributions:}
% 1) We propose a Mamba-based myocardium motion tracking network MCM, which effectively captures long-range dependencies in cardiac motion while maintaining linear complexity. 2) We introduce a sequence-based input strategy and a novel bi-directional Mamba scanning mechanism designed for sequences to address the problem of temporal consistency in motion tracking. 3) We design a bidirectional motion fusion module to further enhance the temporal coherence of the estimated motion field. 4) Comprehensive experiments on two public datasets demonstrate that our method achieves state-of-the-art performance, outperforming existing approaches.

\noindent\textbf{Related Works:} 
Deformable image registration methods have been widely applied to cardiac motion tracking, where traditional techniques have demonstrated their efficacy~\cite{Rueckert1999,Vercauteren2007,Craene2012}. For instance, Vercauteren et al.~\cite{Vercauteren2007} introduced the non-parametric diffeomorphic approach based on the demons algorithm~\cite{Thirion1998}, which has been effectively used for cardiac motion tracking~\cite{QIN2023102682}. More recently, deep learning-based image registration methods have gained increased attention. Balakrishnan et al.~\cite{Balakrishnan2019} proposed VoxelMorph, which employs a U-Net architecture for registration and has been extended to cardiac motion estimation~\cite{Meng2022_mulvimotion}. Chen et al.~\cite{CHEN2022} developed TransMorph, utilizing vision transformers to capture long-range spatial relationships. Building on neighborhood attention, Wang et al.~\cite{Wang2023} introduced ModeT to further improve the interpretability and consistency of deformation estimation. Lately, inspired by State Space Models (SSM)~\cite{5311910}, Mamba~\cite{gu2024mamba} has been developed to address the limitations of modeling lengthy sequences, and it has been explored in various medical image analysis tasks~\cite{UMamba,Yang2024,Liu2024,Yang_MambaMIL2024,Wang_TriPlane_MICCAI2024,guo2024mambamorph,wang2024vmambamorph}.
% The main idea is to effectively capture long-range dependencies by selecting scanning mechanisms and implementing hardware-aware algorithms. As Mamba becomes a new alternative to conventional CNNs and transformers, it has been explored in various medical image analysis tasks~\cite{UMamba,Yang2024,Liu2024,Yang_MambaMIL2024,Wang_TriPlane_MICCAI2024}, including medical image registration~\cite{guo2024mambamorph,wang2024vmambamorph}. 
% Different from existing deep learning-based cardiac motion estimation methods, which consider isolated frame pairs, our method incorporates sequences of target images to enhance the smoothness and temporal consistency of the estimated motion fields.
In contrast to existing cardiac motion estimation methods that rely on isolated frame pairs, our method leverages Mamba to process sequences of target images, enhancing the temporal consistency of the estimated motion fields.
% Moreover, unlike Vision Mamba~\cite{2024visionmamba}, which applies bidirectional scanning on single 2D images, our proposed method employs a bidirectional scanning mechanism specifically designed for sequential cardiac image frames to capture spatiotemporal dependencies, thereby improving the accuracy of myocardial motion tracking.
Our proposed bi-directional scanning mechanism is tailored to sequential cardiac image frames, going beyond prior methods such as Vision Mamba~\cite{2024visionmamba}, which apply bidirectional scanning only to single 2D images.

\section{Method}
\begin{figure}[t]
    \centering
    \includegraphics[width=0.91\textwidth]{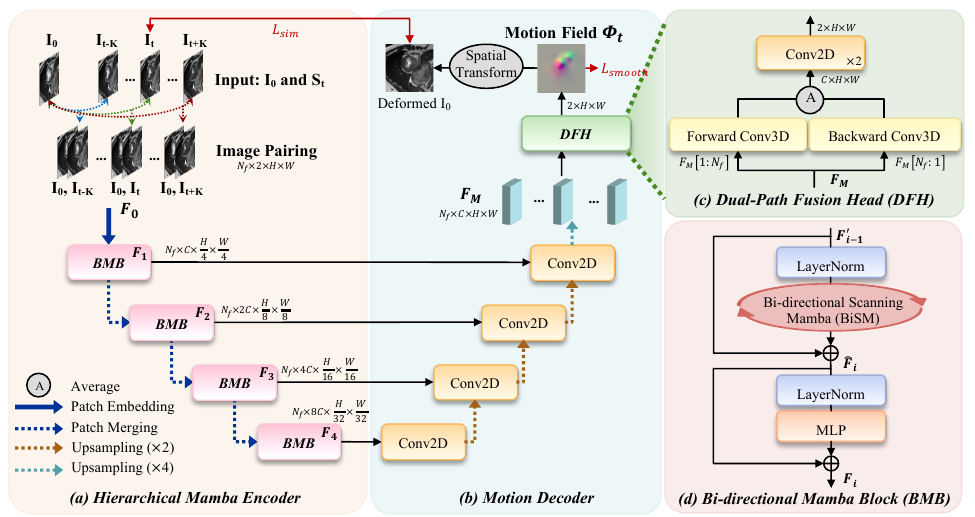}
    \caption{An overview of the proposed method: (a) The Hierarchical Mamba encoder pairs the reference image with the target image sequence and learns deformation features at different scales; (b) The motion decoder combines the learned deformation features at various scales and predicts the motion field $\Phi_t$ via a Dual-Path Fusion Head (DFH); (c) The detailed network architecture of DFH; (d) The detailed network architecture of Bi-directional Mamba Block (BMB).}
    \label{fig:framework}
\end{figure}
Our goal is to estimate LV myocardial motion from 2D short-axis (SAX) CMR images. Our task is formulated as follows: Let $I_0$ be the SAX image of the end-diastole (ED) frame, \emph{i.e.,} reference frame, and $I_t$ be the image of the $t$-th frame ( $0\leqslant t\leqslant T-1$), \emph{i.e.,} target frame. $T$ is the number of frames in the cardiac cycle. We aim to estimate a motion field $\Phi_t$ between ED and $t$-th frame.

The schematic architecture of the proposed method is shown in Fig.~\ref{fig:framework}.
Our method leverages sequences of target images for motion estimation, using the ED frame and $K$ neighboring frames around the $t$-th frame to estimate $\Phi_t$. We denote the sequence of target frames $\mathcal{S}_t = \{I_{t-K}, \ldots, I_t, \ldots, I_{t+K}\}$.
The method comprises two main components. First, a hierarchical Mamba encoder pairs the input images (reference and targets) and learns deformation features at multiple scales via bi-directional Mamba blocks. Within each Mamba block, the proposed bi-directional scanning mechanism is used to integrate spatiotemporal information from both forward and backward directions, facilitating the estimation of plausible deformation. Second, a motion decoder combines the learned deformation features across all scales to predict the motion field $\Phi_t$. Particularly, a dual-path fusion head is developed to strengthen the temporal consistency of $\Phi_t$.
% First, the input images are paired and bi-directionally scanned to explicitly capture spatial coherence within frames and temporal coherence between frames. Second, a Mamba-backboned encoder extracts both coarse and fine spatiotemporal features, facilitating the investigation of correlations between spatial and temporal dependencies across various scales. Finally, the motion estimation head iteratively aggregates the multi-scale features of the entire sequence through a hierarchical decoder, and then integrates the motion cues from the forward and reverse directions through a bi-directional motion fusion module to produce the motion field $\Delta \Phi_t$ of the target frame.

% \subsection{Bi-direction spatiotemporal scanning}

\begin{figure}[t]
    \centering
    \includegraphics[width=1\textwidth]{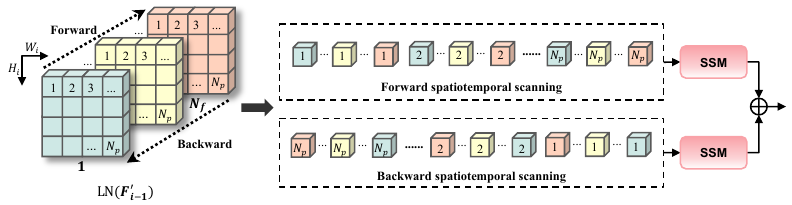}
    \caption{The proposed bi-directional scanning Mamba (BiSM), including forward and backward spatiotemporal scanning. \(\text{LN}(F_{i-1}^{\prime})\) is \(F_{i-1}^{\prime}\) after layer normalization. $N_p = H_i \times W_i$ is the total number of spatial positions.}
    \label{fig:scan}
\end{figure}

\subsection{Hierarchical Mamba Encoder}

The hierarchical Mamba encoder learns multi-scale deformation features $F_i$ from the input images. Specifically, the input images $I_0$ and $\mathcal{S}_t$ are paired into an input sequence, which is then forwarded to the hierarchical Mamba blocks to learn $F_i$. Within each Mamba block, a bi-directional scanning mechanism is developed to learn spatiotemporal information from the input sequence.
% The encoder hierarchically extracts multi-scale spatiotemporal features, progressively downsampling the input while modeling spatial and temporal dependencies. As shown in Fig.~\ref{fig:framework}(a), the input first passes through an overlapping patch embedding module, which projects the paired images into compact feature representations before feeding them into Bi-Mamba Blocks.
\subsubsection{Image pairing:}

% In this part, we combine $I_0$ with the target images sequence $S_t$, \emph{i.e.,} $I_t$ and its $K$ neighboring frames, to form the paired images input sequence. 
In this part, we pair $I_0$ with $S_t$ to form the input sequence $F_0$. 
% 
% Formally, we define $F_0 = \{(I_0, I_{t-K}), \ldots, (I_0, I_t), \ldots, (I_0, I_{t+K})\}$ of length $N_f = 2K + 1$, where each frame from $\mathcal{S}_t$ is paired with the same input image $I_0$. Each pair in $F_0$ has a shape of $[2, H, W]$, where $H$ and $W$ are the resolution of the input images.
In detail, each frame from $\mathcal{S}_t$ is paired with the same input image $I_0$ and $F_0 = \{(I_0, I_{t-K}), \ldots, (I_0, I_t), \ldots, (I_0, I_{t+K})\}$. Each pair in $F_0$ has a shape of $[2, H, W]$, where $H$ and $W$ are the height and width of the input images, and $F_0$ has a length of $N_f = 2K + 1$.
% 
% In detail, $I_0$ is replicated $N_f$ times, where $ N_f = 2K + 1$ represents the number of frames in the sequence. Each replicated $I_0$ is paired with one frame $\{I_{t-K+n}|n\in[0,N_f)\}$ in the sequence, resulting in $N_f$ paired images where each pair has a shape of $[1, 2, H, W]$. By concatenating all $N_f$ pairs, we obtain the paired images $F_0 \in \mathbb{R}^{N_f \times 2 \times H \times W}$ for the following hierarchical Mamba blocks.
% 
Note that if $t<K$ or $t>T-K-1$, we use the nearest available frame for padding \emph{e.g.} for $t=1, K=2$, $\mathcal{S}_1 = \{I_0, I_0, I_1, I_2, I_3\}$.

\subsubsection{Hierarchical Mamba blocks:}

From $F_0$, hierarchical bi-directional Mamba blocks (BMBs) are utilized to learn deformation features $F_i \in \mathbb{R}^{N_f \times C_i \times H_i \times W_i}$ at multiple scales. Here, $C_i$, $H_i$, and $W_i$ represent the number of channels, height, and width of $F_i$ at the $i$-th level. 
% Following the patch embedding process, several consecutive stages of patch merging and Bi-directional Mamba Blocks are applied on $F_i$. As shown in Fig.~\ref{fig:framework}(d), the $i$-th Bi-directional Mamba Block process the deformation features as:
Patch embedding or patch merging~\cite{CHEN2022} are used to downsample $F_i$ between two BMBs. As shown in Fig.~\ref{fig:framework}(d), the $i$-th BMB process the deformation features as:
% \begin{equation}
% F_i = \text{MLP}(\text{LN}({BSM}(\text{LN}(F_{i-1}^{\prime})) + F_{i-1}^{\prime})) + BSM(\text{LN}(F_{i-1}^{\prime})) + F_{i-1}^{\prime},i\in[1,4].
% \end{equation}
\begin{align}
    \hat{F}_i &= \text{BiSM}(\text{LN}(F_{i-1}^{\prime})) + F_{i-1}^{\prime},\\
    F_i &= \text{MLP}(\text{LN}(\hat{F}_i)) + \hat{F}_i, \quad i\in[1,4].
\end{align}
Here, $\text{BiSM}(\cdot)$ represents the bi-directional scanning Mamba, which will be discussed next. $F_{i-1}^{\prime}$ is the $F_{i-1}$ after the patch embedding or patch merging. LN is the layer normalization and MLP is a multi-layer perceptron.
% to enrich the feature representation.

% The core mechanism of the Mamba Layer leverages a structured state-space formulation, inspired by Structured State-Space Sequence (S4) models~\cite{gu2022efficiently}. This formulation enables efficient long-range dependency modeling by maintaining a latent state that evolves over time. Specifically, the recurrence follows:
% \begin{equation}
%     \mathbf{h}_{k} = \mathbf{A}_d \mathbf{h}_{k-1} + \mathbf{B}_d \mathbf{x}_k,
% \end{equation}
% \begin{equation}
%     \mathbf{y}_k = \mathbf{C}_d \mathbf{h}_k,
% \end{equation}
% where $\mathbf{h}_k$ represents the latent state at step $k$, and $\mathbf{A}_d, \mathbf{B}_d, \mathbf{C}_d$ are structured transition matrices optimized for long-range dependency modeling. Mamba further enhances S4 by dynamically adjusting state transitions and leveraging a hardware-optimized implementation, improving both efficiency and scalability~\cite{gu2024mamba}.

% The stacked Mamba Blocks are arranged hierarchically to form a multi-resolution encoder. As the network progresses, the spatial resolution decreases while the number of channels increases, transitioning from high-resolution local motion details to lower-resolution global motion patterns. This progressive encoding ensures both fine and coarse cardiac motion are captured effectively, providing a rich feature representation for the motion estimation head.

\subsubsection{Bi-directional scanning Mamba (BiSM):}

% In each bi-directional Mamba block, a bi-directional scanning Mamba (BSM) layer is developed to integrate spatiotemporal information at the $i$-th level.Fig.~\ref{fig:scan} shows the proposed scanning mechanism. Our scanning mechanism explicitly considers spatial coherence over each frame and temporal coherence across the paired-frame sequence. 
% Parallel SSMs are leveraged to establish the intra- and inter-frame long-range dependencies. The results from both forward and backward scanning directions are fused by summation, ensuring a unified spatiotemporal representation that enables plausible deformation features.

Each BMB incorporates a BiSM to integrate spatiotemporal information at level $i$, as illustrated in Fig.~\ref{fig:scan}. 
% To prepare for temporal modeling, $\text{LN}(F_{i-1}^{\prime})$ is reshaped to \(N_p\) spatial positions, where \(N_p\) denotes the number of positions per frame. 
To prepare for temporal modeling, $\text{LN}(F_{i-1}^{\prime})$ is split into \(N_p\) spatial positions per frame, where $N_p=H_i\times W_i$ is the number of positions.
These positions are then temporally ordered in both forward and backward directions and fed into two parallel SSMs. Each SSM captures temporal dynamics by recursively updating hidden states through learned linear recurrence. The outputs from both directions are summed to form a unified spatiotemporal representation, enabling smooth and consistent deformation estimation.

\subsection{Motion Decoder}

The proposed motion decoder estimates the motion field $\Phi_t$ by progressively integrating multi-scale deformation features $F_i$. It consists of a progressive upsampling pathway and a dual-path fusion head (shown in Fig.~\ref{fig:framework}(b)). 
The progressive upsampling pathway $\text{PUP}(\cdot)$ fuses deformation features $F_i,i \in [1, 4]$ via multiple upsampling and convolutional layers and estimates the motion feature $F_M\in \mathbb{R}^{N_f \times C \times H \times W}$ that represents the deformation of the sequential images:
\begin{equation}
    F_M = \text{PUP}(\left\{ F_i \mid  i \in \left[1,4 \right] \right\} ).
\end{equation}
%For decoder without temporal modeling, we reshape $(Batch, N_f, C, H, W)$ to $(Batch \times N_f, C, H, W)$ to match the input format.
%In practice, we reshape $(\text{Batch}, N_f, C, H, W)$ to $(\text{Batch} \times N_f, C, H, W)$ in our experiments, as the decoder processes frames independently.

% 
% The proposed motion decoder estimates the motion field $\Phi_t$ by progressively integrating multi-scale deformation features $F_i$. It consists of a progressive upsampling pathway  and a dual-path fusion head (shown in Fig.~\ref{fig:framework}(b)). 
% The progressive upsampling pathway fuses deformation features $F_i, i \in [1, 4])$ via a Multi-Scale Fully Convolutional Network $\text{MSFCN}(\cdot)$ and predicts the motion feature $F_M\in \mathbb{R}^{N_f \times C \times H \times W}$ that represents the deformation of all the sequential images:
% \begin{equation}
%     F_M = \text{MSFCN}(\left\{ F_i \mid  i \in \left[1,4 \right] \right\} ).
% \end{equation}
% % 

% gradually restores high-resolution motion details by iteratively refining the feature maps. At each stage, the feature map undergoes spatial upsampling, followed by fusion with the corresponding encoder features through skip connections, ensuring a detailed and accurate motion representation.
To further enforce temporal coherence across frames, a dual-path fusion head $\text{DFH}(\cdot)$ is introduced to estimate $\Phi_t\in\mathbb{R}^{2 \times H \times W}$ from $F_M$. The architecture of $\text{DFH}(\cdot)$ is shown in Fig.~\ref{fig:framework}(c). Specifically, $F_M$ is simultaneously passed in the forward direction (from 1 to $N_f$) and the backward direction (from $N_f$ to 1) via 3D convolutional layers operating along $N_f$, $H$ and $W$. Subsequently, the results from both paths are averaged, and then passed to 2D convolutional layers to estimate $\Phi_t$:
% \begin{equation}
%     \Phi_t = DFH(F_M)
% \end{equation}
\begin{align}
    \overline{F}_M &= \frac{1}{2} \left( \text{Conv3D}_{\text{fwd}}(F_M[1:N_f]) + \text{Conv3D}_{\text{bwd}}(F_M[N_f:1]) \right),\\
    \Phi_t &= \text{DFH}(F_M) = \text{Conv2Ds}\left( \overline{F}_M \right).
\end{align}

\subsection{Optimization}

Our model is an end-to-end trainable framework, and the overall objective is a linear combination of two loss terms:

\begin{equation}\label{loss}
    \mathcal{L} = \underbrace{\frac{1}{|\Omega|} \sum_{p \in \Omega} \| I_t(p) - I_0\circ\Phi_t(p) \|^2}_{\ \mathcal{L}_{sim}} 
    + \lambda \underbrace{\sum_{p \in \Omega} \|\nabla \Phi_t (p)\|^2}_{ \ \mathcal{L}_{smooth}},
\end{equation}
where $\lambda$ is the weight of the regularization term, $p$ is a pixel in the image domain $\Omega$ and $|\Omega|$ is the total number of pixels. The similarity loss $\mathcal{L}_{sim}$ is defined by the mean squared error while $\mathcal{L}_{smooth}$ is the smoothness regularization.

\section{Experiments}

\noindent\textbf{Dataset: } We evaluate the proposed method on two publicly available cine CMR datasets: ACDC~\cite{Bernard2018} and M\&Ms~\cite{Campello2021}. Both datasets provide a series of short-axis (SAX) slices covering the left ventricle (LV) from the base to the apex. All image slices are resampled to a resolution of $1.5 \times 1.5$ mm, center-cropped to $128 \times 128$ pixels and normalized to $[0,1]$. The ACDC dataset is divided into 80/20/50 for training, validation, and testing, respectively, while the M\&Ms dataset follows a 150/34/136 split. 

\noindent\textbf{Evaluation metrics: } Quantitative evaluation is performed using three commonly used metrics: the Dice coefficient to assess motion tracking accuracy, the percentage of negative Jacobian determinant values ($|J|_{<0}\%$) to evaluate diffeomorphism, and the mean absolute difference between $|J|$ and 1 (\emph{i.e.,}$||J|-1|$) to measure volume preservation. A higher Dice score indicates better tracking performance, while lower $|J|_{<0}\%$ and $||J|-1|$ values indicate improved diffeomorphic properties and volume consistency, respectively.

\noindent\textbf{Implementation: } The proposed model is implemented in PyTorch and trained on an NVIDIA A100-SXM4 GPU with 40GB of memory. Network optimization is performed using the Adam optimizer with a learning rate of $10^{-4}$. The model is trained for 200 epochs on both datasets with a batch size of 32. The hyper-parameter in Eq.~\ref{loss} is set to $\lambda = 0.05$ for both datasets. We estimate the motion fields for all frames in the cardiac cycle.

\begin{table}[t]
    \centering
    \caption{Quantitative comparison of other cardiac motion tracking methods. The results are reported as "mean (standard deviation)". $\uparrow$ indicates the higher value the better while $\downarrow$ indicates the lower value the better. Best results in bold.}
    \label{table:comparison}
    \resizebox{\textwidth}{!}{%
    \begin{tabular}{p{0.3cm}lcccrcccrccc}
    \toprule
        ~~~ & 
        ~~~ & 
        \multicolumn{3}{c}{\textbf{Basal}} & \hspace{0.25cm} &
        \multicolumn{3}{c}{\textbf{Mid-ventricle}} & \hspace{0.25cm} &
        \multicolumn{3}{c}{\textbf{Apical}}\\
        ~~~ & 
        ~~~ & 
        \small{Dice\%$\uparrow$} &
        \small{$|J|_{<0}\%$$\downarrow$} &
        \small{$||J|-1|$$\downarrow$} & &
        \small{Dice\%$\uparrow$} &
        \small{$|J|_{<0}\%$$\downarrow$} &
        \small{$||J|-1|$$\downarrow$} & &
        \small{Dice\%$\uparrow$} &
        \small{$|J|_{<0}\%$$\downarrow$} &
        \small{$||J|-1|$$\downarrow$} \\
        \cmidrule(lr){1-5}\cmidrule(lr){7-9} \cmidrule(lr){11-13}
        % ~~~ & 
        % \multicolumn{9}{c}{\textbf{ACDC dataset}} \\
        % \cmidrule{2-4}\cmidrule{5-7} \cmidrule{8-10}
        \multirow{5}{*}{\rotatebox[origin=c]{90}{\textbf{ACDC}}} &
        dD~\cite{Vercauteren2007} &
        78.9(10.7) & 0.35(0.30) & 0.29(0.05) & & % Apical
        80.9(7.2) & 0.36(0.24) & 0.30(0.05) & & % Mid-ventricle
        78.6(8.7) & 0.28(0.19) & 0.29(0.05) \\ % Basal
        ~~~ &
        VM~\cite{Balakrishnan2019} &
        81.5(6.9) & 0.27(0.42) & 0.25(0.16) & & % Apical
        81.0(7.1) & 0.08(0.14) & 0.27(0.13) & & % Mid-ventricle
        79.1(8.5) & 0.03(0.09) & 0.28(0.13) \\ % Basal
        ~~~ &
        TM~\cite{CHEN2022} &
        82.6(7.3) & 0.28(0.40) & 0.19(0.07) & & % Apical
        83.7(4.9) & 0.05(0.09) & 0.19(0.07) & & % Mid-ventricle
        82.4(5.9) & 0.02(0.05) & 0.19(0.09) \\ % Basal
        ~~~ &
        MM~\cite{guo2024mambamorph} &
        82.2(6.8) & 0.33(0.48) & 0.19(0.07) & & % Apical
        83.7(5.3) & 0.05(0.09) & 0.19(0.07) & & % Mid-ventricle
        82.3(5.8) & 0.05(0.10) & 0.20(0.08) \\ % Basal
        ~~~ &
        Ours  &
        \textbf{83.4(7.1)} & \textbf{0.14(0.31)} & \textbf{0.17(0.06)} & & % Apical
        \textbf{84.6(4.9)} & \textbf{0.02(0.04)} & \textbf{0.18(0.06)} & & % Mid-ventricle
        \textbf{82.8(5.5)} & \textbf{0.01(0.02)} & \textbf{0.17(0.06)} \\ % Basal
        \cmidrule(lr){1-5}\cmidrule(lr){7-9} \cmidrule(lr){11-13}
        \multirow{5}{*}{\rotatebox[origin=c]{90}{\textbf{M\&Ms}}} &
        dD~\cite{Vercauteren2007} &
        75.7(11.3) & 0.26(0.21) & 0.30(0.06) & & % Apical
        78.1(8.9) & 0.29(0.22) & 0.27(0.05) & & % Mid-ventricle
        73.4(13.0) & 0.24(0.20) & 0.30(0.07) \\ % Basal
        ~~~ &
        VM~\cite{Balakrishnan2019} &
        74.6(12.5) & 0.09(0.17) & 0.30(0.14) & & % Apical
        79.5(9.8) & 0.21(0.37) & 0.29(0.14) & & % Mid-ventricle
        74.6(12.3) & 0.29(0.38) & 0.27(0.12) \\ % Basal
        ~~~ &
        TM~\cite{CHEN2022} &
        79.1(8.5) & 0.11(0.24) & 0.20(0.08) & & % Apical
        82.0(6.0) & 0.23(0.40) & 0.20(0.07) & & % Mid-ventricle
        76.4(11.7) & 0.26(0.51) & 0.20(0.09) \\ % Basal
        ~~~ &
        MM~\cite{guo2024mambamorph} &
        78.7(8.9) & 0.08(0.17) & \textbf{0.19(0.07)} & & % Apical
        82.2(6.2) & 0.20(0.35) & 0.19(0.07) & & % Mid-ventricle
        76.1(12.0) & 0.22(0.46) & 0.20(0.09) \\ % Basal
        ~~~ &
        Ours  &
        \textbf{79.9(8.4)} & \textbf{0.03(0.09)} & \textbf{0.19(0.07)} & & % Apical
        \textbf{83.6(6.2)} & \textbf{0.12(0.29)} & \textbf{0.18(0.06)} & & % Mid-ventricle
        \textbf{77.6(11.5)} & \textbf{0.15(0.40)} & \textbf{0.19(0.08)} \\ % Basal
        \bottomrule
    \end{tabular}
    }
\end{table}

\begin{figure}[t]
    \centering
    \includegraphics[width=1\textwidth]{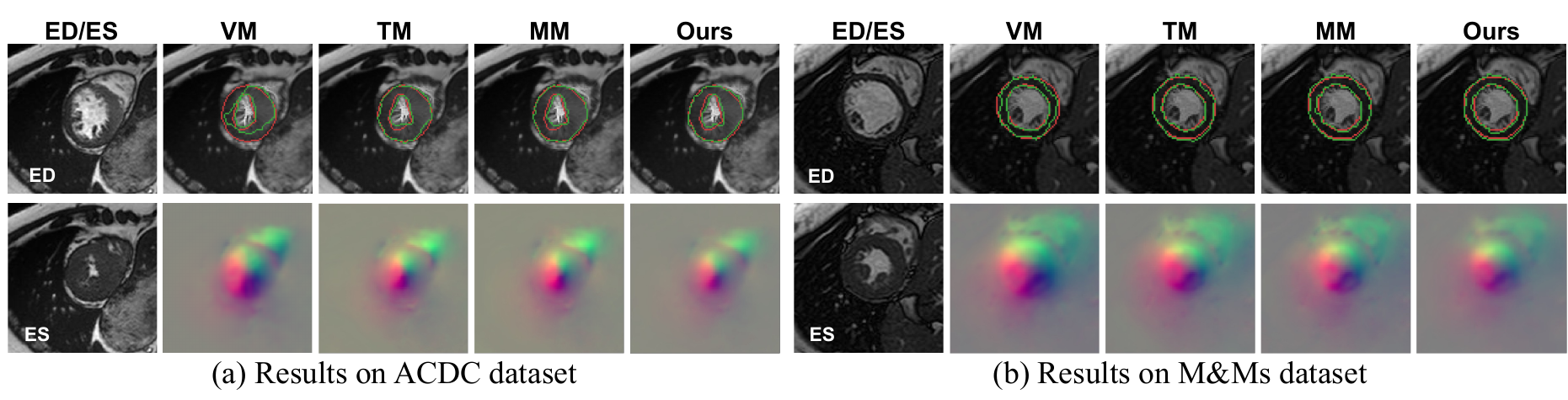}
    \caption{Motion tracking results using proposed method and baselines. We warp the ED segmentation to the ES frame. The top row shows the deformed ED myocardial contour (green) vs. the ground truth ES myocardial contour (red). The bottom row shows the estimated motion fields.
    }
    \label{fig:compare}
\end{figure}

\noindent\textbf{Comparison study: } The proposed method is compared to one conventional cardiac motion tracking method, dDemons (dD)~\cite{Vercauteren2007} and three the state-of-the-art deep learning-based methods, including VoxelMorph (VM)~\cite{Balakrishnan2019}, TransMorph (TM)~\cite{CHEN2022} and MambaMorph (MM)~\cite{guo2024mambamorph}. All methods are implemented using their officially released code, with optimal parameters tuned on the validation sets. 
Quantitative comparisons were performed on three representative short-axis slices: basal, mid-ventricular and apical slices, corresponding to 25$\%$, 50$\%$ and 75$\%$ of the LV length, respectively. We choose $K=2$, and thus have the input sequential images with $N_f = 5$ frames. In this experiment, we estimate the motion field between the ED frame and the end-systolic (ES) frame and warp the ED frame segmentation to the ES frame, and compute evaluation metrics by comparing the wrapped segmentation with the ground truth ES segmentation. 
Table~\ref{table:comparison} shows that the proposed method outperforms all baseline methods, demonstrating the effectiveness of the proposed method for estimating motion fields. In addition, the proposed method achieves the best performance on $|J|_{<0}\%$ and $||J|-1|$ for all three slices, indicating that the proposed method is more capable of estimating smooth motion fields and preserving the volume of the myocardial wall during cardiac motion tracking.
We further qualitatively compare the proposed method with baselines. Fig.~\ref{fig:compare} shows that the motion field generated by the proposed method performs best in warping the ED segmentation to the ES frame, and it is the smoothest. This demonstrates that our method is able to estimate smooth and consistent motion fields.

\begin{figure}[t]
    \centering
    \includegraphics[width=0.7\textwidth]{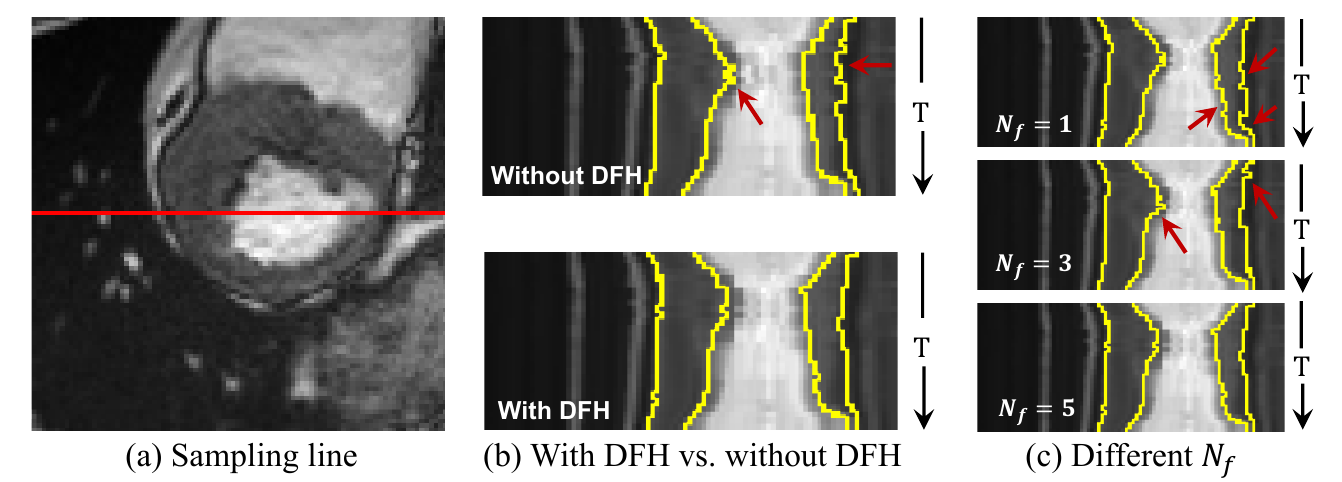}
    \caption{Temporal consistency across the cardiac cycle. The red line in (a) denotes the temporal axis for (b) and (c).}
    \label{fig:head}
\end{figure}

\begin{figure}[tbp]
    \centering
    \begin{minipage}[c]{0.5\textwidth}
        \centering
        \includegraphics[width=\textwidth]{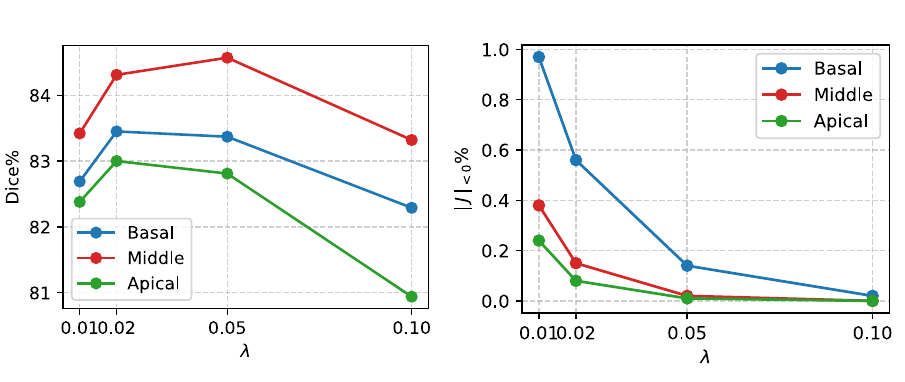} 
        \caption{Motion estimation with different values of $\lambda$.}
        \label{fig:lambda}
    \end{minipage} \hspace{0.1cm}
    \begin{minipage}[c]{0.45\textwidth}
        \centering
        \captionsetup{type=table} % Change caption type to table
        \caption{GPU VRAM and inference time of comparison methods.}
        \scriptsize
        \begin{tabular}{lcc}
            \toprule
            ~~~ & VRAM (GB) & time (ms) \\ \hline
            VM~\cite{Vercauteren2007} & 1.5 & \: 7.9 \\
            TM~\cite{Balakrishnan2019} & 3.6 & 14.9 \\
            MM~\cite{guo2024mambamorph} & 2.7 & 22.9 \\
            Ours ($N_f$=1) & 3.2 & 16.3 \\
            Ours ($N_f$=3) & 7.8 & 16.5 \\
            Ours ($N_f$=5) & 12.4 & 17.1 \\
            \bottomrule
        \end{tabular}
        \label{tab:efficiency}
    \end{minipage}
\end{figure}

\begin{table}[tp]
    \centering
    \caption{Motion estimation without BMBs and with BMBs using different scanning strategies.}
    \label{table:ablation}
    \resizebox{0.9\textwidth}{!}{%
    \begin{tabular}{p{0.3cm}lccrccrcc}
    \toprule
        ~~~ & 
        ~~~ & 
        % \multicolumn{3}{c}{\textbf{Basal}} & 
        % \multicolumn{3}{c}{\textbf{Mid-ventricle}} & 
        % \multicolumn{3}{c}{\textbf{Apical}}\\
        \multicolumn{2}{c}{\textbf{Basal}} & \hspace{0.3cm} &
        \multicolumn{2}{c}{\textbf{Mid-ventricle}} & \hspace{0.3cm} &
        \multicolumn{2}{c}{\textbf{Apical}}\\
        ~~~ & 
        ~~~ & 
        \small{Dice\%$\uparrow$} &
        \small{$|J|_{<0}\%$$\downarrow$} & &
        % \small{$||J|-1|$$\downarrow$} &
        \small{Dice\%$\uparrow$} &
        \small{$|J|_{<0}\%$$\downarrow$} & &
        % \small{$||J|-1|$$\downarrow$} &
        \small{Dice\%$\uparrow$} &
        \small{$|J|_{<0}\%$$\downarrow$} \\
        % \small{$||J|-1|$$\downarrow$} \\
        % \cmidrule{3-5}\cmidrule{6-8} \cmidrule{9-11}
        \cmidrule{1-4}\cmidrule{6-7} \cmidrule{9-10}
        % ~~~ & 
        % \multicolumn{9}{c}{\textbf{ACDC dataset}} \\
        % \cmidrule{2-4}\cmidrule{5-7} \cmidrule{8-10}
        \multirow{4}{*}{\rotatebox[origin=c]{90}{\textbf{ACDC}}} &
        Without BMBs &
        81.8(7.8) & 0.15(0.29) & & % 0.17(0.06) & % Apical
        82.4(4.7) & \textbf{0.01(0.02)} & & % 0.17(0.06) & % Mid-ventricle
        81.4(6.1) & \textbf{0.01(0.02)} \\ % & 0.18(0.06) \\ % Basal
        ~~~ &
        BMBs+forward scanning &
        83.1(6.6) & 0.18(0.31) & & % 0.17(0.06) & % Apical
        84.0(4.7) & 0.02(0.04) & & % 0.17(0.06) & % Mid-ventricle
        82.2(5.4) & \textbf{0.01(0.02)} \\ %  & 0.18(0.06) \\ % Basal
        ~~~ &
        BMBs+backward scanning&
        82.8(6.8) & 0.15(0.28) & & % 0.17(0.06) & % Apical
        83.4(4.8) & \textbf{0.01(0.02)} & & % 0.17(0.06) & % Mid-ventricle
        82.0(5.3) & \textbf{0.01(0.02)} \\ % & 0.18(0.06) \\ % Basal
        ~~~ &
        BMBs+BiSM (ours) &
        \textbf{83.4(7.1)} & \textbf{0.14(0.31)} & & % \textbf{0.17(0.06)} & % Apical
        \textbf{84.6(4.9)} & 0.02(0.04) & & % \textbf{0.17(0.06)} & % Mid-ventricle
        \textbf{82.8(5.5)} & \textbf{0.01(0.02)} \\ % & \textbf{0.17(0.06)} \\ % Basal\
        % \midrule
        % \multirow{3}{*}{\rotatebox[origin=c]{90}{\textbf{M\&Ms}}} &
        % Forward &
        % 78.8(8.1) & 0.04(0.12) & % 0.19(0.07) & % Apical
        % 83.1(5.9) & \textbf{0.12(0.28)} & % \textbf{0.17(0.06)} & % Mid-ventricle
        % 76.8(11.1) & \textbf{0.15(0.37)} \\ % & \textbf{0.18(0.08)} \\ % Basal
        % ~~~ &
        % Backward &
        % 78.9(8.1) & \textbf{0.03(0.09)} & % 0.19(0.07) & % Apical
        % 83.2(6.1) & 0.13(0.30) & % \textbf{0.17(0.06)} & % Mid-ventricle
        % 77.2(11.1) & 0.16(0.38) \\ % & 0.19(0.08) \\ % Basal
        % ~~~ &
        % Bi-directional &
        % \textbf{79.9(8.4)} & \textbf{0.03(0.09)} & % \textbf{0.19(0.07)} & % Apical
        % \textbf{83.6(6.2)} & \textbf{0.12(0.29)} & % 0.18(0.06) & % Mid-ventricle
        % \textbf{77.6(11.5)} & \textbf{0.15(0.40)} \\ % & 0.19(0.08) \\ % Basal
        \bottomrule
    \end{tabular}
    }
\end{table}
\noindent\textbf{Ablation study: } On the ACDC dataset, we explore the importance of BMBs, BiSM and DFH, as well as the effects of hyper-parameters. 
% Table~\ref{table:ablation} shows that bi-directional scanning performs best, illustrating the importance of the proposed approach. 
% Table~\ref{table:ablation} compares different scanning strategies. The bi-directional scanning performs best, while removing BMBs leads to the worst performance, confirming the importance of explicit temporal modeling.
Table~\ref{table:ablation} shows that our method, incorporating both BMBs and BiSM, achieves the best performance, while removing BMBs results in the poorest performance. This indicates that the performance gain stems from our proposed approach rather than merely from an increased number of input frames.
Fig.~\ref{fig:head} (b) shows that the motion estimation with DFH achieves better temporal consistency across the cardiac cycle, supporting the importance of the proposed DFH. Fig.~\ref{fig:head} (c) shows the temporal consistency variations when using different target sequence lengths. We observe that using more neighboring frames achieves better temporal smoothness. Fig.~\ref{fig:lambda} presents the quantitative metrics with various $\lambda$ in Eq.~\ref{loss}. We observe that a strong constraint on motion field smoothness may sacrifice motion estimation accuracy.

\noindent\textbf{Computational cost: }
% We evaluate model efficiency using GPU training memory (i.e., VRAM) and inference time. As shown in Table~\ref{tab:efficiency}, the proposed method maintains a computational cost comparable to the baselines, demonstrating its ability to perform efficient motion estimation using sequential images without additional computational overhead.
We evaluate model efficiency using GPU training memory (i.e., VRAM) and inference time. As shown in Table~\ref{tab:efficiency}, while VRAM usage increases with larger $N_f$ due to buffering multiple frames, the inference time remains comparable to baselines, indicating efficient use of sequential images without significant overhead.

\noindent\textbf{Discussion: } 
We quantitatively evaluated the performance of our model for ED to ES motion estimation.
This is because ground truth segmentation are only available for the ED and ES frames in both datasets. Motion fields were estimated on three representative SAX slices across the LV, consistent with existing motion estimation studies~\cite{QIN2023102682}. Our method also facilitates motion estimation using all slices, at the cost of increased GPU memory usage and longer training time.  
% Importantly, the observed performance improvement is not simply due to inputting more frames, but rather arises from the proposed bi-directional scanning mechanism within Mamba, which explicitly leverages temporal coherence across sequential frames to improve myocardial motion tracking.
% As the core contribution of our bi-directional scanning Mamba is explicitly designed to enhance temporal consistency, we observed only slight improvements in quantitative metrics when increasing $N_f$ from 1 to 5 (e.g., Dice: +0.5\%). However, as clearly shown in Fig.~\ref{fig:head}, a larger $N_f$ significantly enhances the temporal smoothness and consistency of the estimated motion fields, underscoring the effectiveness of our proposed method in capturing temporal dependencies.
As our bi-directional Mamba is designed to improve temporal consistency, increasing $N_f$ from 1 to 5 yields only modest gains in quantitative metrics (\emph{e.g.}, +0.5\% in Dice) but results in visibly smoother motion fields, as shown in Fig.~\ref{fig:head}.
% While our experiments focus on 2D motion tracking due to the availability of public 2D datasets, the proposed method can easily be adapted to 3D myocardial motion tracking by incorporating 3D convolutional layers in our model.
% Our experiments focus on 2D motion tracking due to the availability of public 2D datasets. Although our method effectively captures deformation fields and temporal coherence, internal myocardial texture changes still pose a challenge common to existing cardiac registration methods. Future work may extend our framework to 3D cardiac motion tracking by incorporating 3D convolutional layers and exploring advanced texture-preserving regularizations.
Our experiments focus on 2D motion tracking due to the use of publicly available 2D datasets. Future work may extend our framework to 3D by integrating 3D convolutions.

% \begin{figure}[t]
%     \centering
%     \includegraphics[width=0.6\textwidth]{figures/nf.pdf}
%     \caption{(a)Sampling line. (b)Myocardial motion throughout the cycle at the sampling line position with different $N_f$.}
%     \label{fig:nf}
% \end{figure}

% \begin{figure}[t]
%     \centering
%     \includegraphics[width=1\textwidth]{figures/lambda.pdf}
%     \caption{Dice\%, $|J|_{<0}\%$, $||J|-1|$ with varied regularization parameter $\lambda$.}
%     \label{fig:lambda}
% \end{figure}

% \begin{table}[t]
%     \centering
%     \caption{Comparison of the number of parameters, GPU training memory, and averaged GPU inference time.}
%     \label{table:efficiency}
%     \resizebox{0.9\textwidth}{!}{%
%     \begin{tabular}{l >{\centering\arraybackslash}p{3.5cm} >{\centering\arraybackslash}p{3.5cm} >{\centering\arraybackslash}p{3.5cm}}
%     \toprule
%         ~~~ & 
%         Parameters (M) & Training memory (MiB) & Inference time (s) \\
%         \midrule
%         VM & 0.1 & 1496 & 0.0079 \\
%         TM & 31.0 & 3604 & 0.0149 \\
%         MM & 3.5 & 2712 & 0.0229 \\
%         Ours & 16.3 & 7842 & 0.0165 \\
%         \bottomrule
%     \end{tabular}
%     }
% \end{table}

\section{Conclusion}
In this paper, we propose an end-to-end trainable, Mamba-based network for myocardial motion tracking. Our method leverages sequential images to achieve smooth and temporally consistent motion estimation while maintaining computational efficiency. Experimental results on two datasets demonstrate that the proposed method outperforms competing methods.

\subsubsection*{Acknowledgments.}The computations described in this research were performed using the Baskerville Tier 2 HPC service. Baskerville was funded by the EPSRC and UKRI through the World Class Labs scheme (EP/T022221/1) and the Digital Research Infrastructure programme (EP/W032244/1) and is operated by Advanced Research Computing at the University of Birmingham.

\begin{comment}  %% removed for anonymized MICCAI 2025 submission.
    
    % The following acknowledgement and disclaimer sections should be removed for the double-blind review process.  
    % If and when your paper is accepted, reinsert the acknowledgement and the disclaimer clause in your final camera-ready version.

\begin{credits}
\subsubsection{\ackname} A bold run-in heading in small font size at the end of the paper is
used for general acknowledgments, for example: This study was funded
by X (grant number Y).

\subsubsection{\discintname}
It is now necessary to declare any competing interests or to specifically
state that the authors have no competing interests. Please place the
statement with a bold run-in heading in small font size beneath the
(optional) acknowledgments\footnote{If EquinOCS, our proceedings submission
system, is used, then the disclaimer can be provided directly in the system.},
for example: The authors have no competing interests to declare that are
relevant to the content of this article. Or: Author A has received research
grants from Company W. Author B has received a speaker honorarium from
Company X and owns stock in Company Y. Author C is a member of committee Z.
\end{credits}

\end{comment}
\bibliographystyle{splncs04}
\bibliography{mybib}
%

\begin{comment}

\end{comment}

\end{document}